\documentclass[conference]{IEEEtran}

\usepackage[top=0.82in, bottom=1.05in, left=0.68in, right=0.68in]{geometry}

\usepackage[T1]{fontenc}
\usepackage{cite}

\ifCLASSINFOpdf
   \usepackage[pdftex]{graphicx}
\else
   \usepackage[dvips]{graphicx}
\fi

\usepackage{amsmath}
\usepackage{bm}

\usepackage{algorithmic}

\usepackage{array}

\usepackage{subfigure}

\usepackage{amsthm}

\usepackage{color}
\usepackage{extarrows}

\usepackage{algorithm}

\usepackage{siunitx}

\usepackage{amssymb}

\newcommand\blfootnote[1]{
  \begingroup
  \renewcommand\thefootnote{}\footnote{#1}
  \addtocounter{footnote}{-1}
  \endgroup
}

\usepackage{multirow}

\begin{document}

\title{
\huge
Reduced-Overhead Channel Estimation and Iterative Detection of FTN Signaling Based on Pilot Superimposition and Spectral Interference Alignment}

\author{
\IEEEauthorblockN{Yuchen~Wu and Shinya~Sugiura$^*$}
\IEEEauthorblockA{Institute of Industrial Science, The University of Tokyo\\
E-mail: \{wuyuchen0031, sugiura\}@g.ecc.u-tokyo.ac.jp} \vspace*{-7mm}
}

\maketitle

\thispagestyle{empty}

\begin{abstract} 
This paper proposes low-overhead and low-complexity channel estimation (CE) of frequency-domain equalization aided faster-than-Nyquist (FTN) signaling. In the proposed CE scheme, the concept of pilot superimposition is employed, where the FTN block is designed to superimpose pilot symbols with information symbols, and thus, no dedicated time and frequency resources nor guard bands are required, resulting in a 50\% reduction of the overhead. Furthermore, interference induced by the pilot superimposition is eliminated by invoking a novel scheme, referred to as spectral interference alignment, where a data-dependent sequence is subtracted from transmitted information symbols. The theoretical mean-square error (MSE) of the proposed CE is derived, which verifies that the MSE is no longer affected by interference due to the pilot superimposition. 
\end{abstract}

\IEEEpeerreviewmaketitle

\section{Introduction}    \label{[SecIntro]}
\blfootnote{Preprint for publication in \textit{IEEE Global Communications Conference (GLOBECOM)}, Taipei, Taiwan, Dec. 2025, pp. 5820-5825, DOI: 10.1109/GLOBECOM59602.2025.11431895.
$\copyright$ 2025 IEEE. Personal use of this material is permitted. Permission from IEEE must be obtained for all other uses, in any current or future media, including reprinting/republishing this material for advertising or promotional purposes, creating new collective works, for resale or redistribution to servers or lists, or reuse of any copyrighted component of this work in other works.}
\IEEEPARstart{F}{aster}-than-Nyquist (FTN) signaling, invented in the 1970s~\cite{MBSTJ75Faster}, has once again become attention-gathering. FTN signaling intentionally breaks the limitation of the Nyquist criterion, enabling FTN-based communication systems to achieve a higher data rate than that of conventional Nyquist-criterion-based systems.
The enhanced data rate of FTN signaling comes at a sacrifice of more challenging signal processing. For communication systems operating at the Nyquist rate, symbols are sent with a time interval of $T_0=1/(2W)$ without any inter-symbol-interference (ISI), where $W$ is the system bandwidth \cite{AR12ProceedFaster}. 
By contrast, FTN symbols are sent with a time interval of $T=\tau T_0$, where $\tau$ is referred to as the packing ratio in the range of $0<\tau<1$ ($\tau=1$ corresponds to the Nyquist scenario).

To guarantee robust and reliable transmission in FTN signaling, high complexity may be imposed on the receiver to deal with ISI. Detection algorithms of FTN signaling can be coarsely categorized into time-domain equalization (TDE) and frequency-domain equalization (FDE). TDEs~\cite{LBZTCOM18reduced} effectively eliminate ISI in the time domain (TD), though the complexity is commonly high. FDEs \cite{SCOML13Freq} eliminate ISI in the frequency domain (FD), usually with the aid of an extra pair of cyclic prefix (CP) and cyclic suffix (CS), which allow the use of discrete Fourier transform (DFT)-based decomposition.

Channel estimation (CE) of FTN systems is another critical issue. Most previous studies on FTN receiver design assume perfect knowledge of channel state information (CSI).
Due to the severe ISI of FTN signaling, there is a trade-off between overhead and complexity for CE. For instance, in \cite{WYGACCESS17A}, a joint CE and detection scheme is proposed, utilizing an FTN-based pilot (FTNP) to achieve high accuracy and low overhead at the cost of high complexity, where affordable complexity can only be achieved with a sufficiently long pilot sequence along with guard bands.
In \cite{ISTWC17Iterative}, CE is carried out with an overhead of $4\nu$, including $2\nu$ for CE and $2\nu$ for FDE, where $\nu$ is the effective ISI length of FTN signaling. In \cite{LGSTVT20Joint}, a joint CE and precoding scheme is designed with an overhead of more than $4\nu$, and hence, achieving accurate CE with low complexity and low overhead is typically challenging.

In order to reduce the pilot overhead in the conventional Nyquist-based transmission, superimposed pilot schemes have been developed~\cite{GMALSP05Channel}, where the pilot symbols are superimposed along with information symbols.
Thus, the pilot symbols do not require dedicated time and frequency resources as well as guard bands. While such superimposition induces interference between pilot and information symbols, it can be eliminated with cancellation techniques. 
There have been several studies adopting superimposed pilot schemes in orthogonal frequency division multiplexing (OFDM)~\cite{HLLTVT09On} and multiple-input multiple-output (MIMO)~\cite{MLXJSAC17On}.
However, the incorporation of the superimposed pilot concept into FTN signaling is an open issue due to the presence of FTN-specific severe ISI.
Moreover, in \cite{GMALSP05Channel}, the scheme, which is referred to as spectral interference alignment (SIA) in this paper, is designed to eliminate interference induced by the pilot superimposition in the FD to guarantee the CE accuracy. To the best of our knowledge, superimposed pilots along with SIA have never been considered for the CE of FTN signaling despite its potential benefits of low overhead and low complexity.

Against the above backcloth, the novel contributions of this paper are as follows.
We propose FD pilot superimposition for efficient CE in FDE-aided FTN signaling, which dispenses with any additional pilot overhead. Additionally, the power of the transmitted signal remained unchanged regardless of the pilot superimposition.
The concept of SIA is invoked to eliminate interference induced by our new pilot superimposition for FTN signaling. More specifically, this allows us to remove the effects of information symbols at specific frequency bins by designing pilot symbols as a data-dependent sequence. 
Furthermore, a low-complexity iterative detection is developed for accurately recovering information symbols.
The theoretical mean-square error (MSE) of the proposed scheme is derived for analyzing the achievable CE performance.
Our performance results demonstrate the bit-error rate (BER) and MSE comparisons between the proposed scheme and the conventional benchmark~\cite{ISTWC17Iterative}.\footnote{{\emph{Notation}:
The expectation and trace are denoted by $\mathbb{E}[\cdot]$ and $\text{Tr}[\cdot]$, respectively. Moreover, $\mathbf{F}_N$ denotes the $N{\times}N$ DFT matrix, whose $(m,n)$-th entry is given by $(1/\sqrt{N})e^{-j2{\pi}mn/N}$ ($m,n=0,1,...,N-1$). $\mathbf{F}_{P,L}$ or $\mathbf{F}_{L,P} (P \geq L)$ denotes the $P{\times}L$ or $L{\times}P$ leading submatrix, respectively, of $\mathbf{F}_{P}\in\mathbb{C}^{P{\times}P}$. $\mathbf{I}_{(\cdot)}$ denotes the identical matrix whose order is decided by the subscript, \emph{e.g.}, $\mathbf{I}_N$ represents an $N{\times}N$ identical matrix. $\mathbf{0}_{(\cdot){\times}(\cdot)}$ denotes the zero matrix whose size is determined by the subscript.}}

\section{System Model of FDE-Aided FTN Signaling} \label{[SecSystemModel]}

\subsection{FTN Signaling in AWGN Channel}
At the transmitter, information bits are firstly modulated onto a block of complex-valued phase shift keying (PSK) or quadrature amplitude modulation (QAM) symbols $s_n~(n=0,1,\cdots,N-1)$ with zero-mean and a variance of $\sigma_s^2$, where $N$ is the block length. The modulated symbols are passed through a root-raised-cosine (RRC) filter with an impulse response (IR) denoted by $q(t)$. 
The pulses are arranged with a time interval of $T=\tau T_0$, which is smaller than the ISI-free Nyquist-based interval. The transmitted FTN signal in the TD $x(t)$ is given by 
$x(t) = \sum_{n=0}^{N-1} s_nq(t-nT)$.
After the signal $x(t)$ is transmitted over the additive white Gaussian noise (AWGN) channel, the received signal is processed by the matched filter, which has the same IR as the RRC filter $q(t)$ at the transmitter, which is represented by 
$y(t) = \sum_{n=0}^{N-1} s_ng(t-nT)+\eta(t)$,
where $g(t)$ and $\eta(t)$ are given by \cite{ISH21ACCESSThe}
$g(t) = \int q(\xi)q^*(\xi-t)\,{\rm d}\xi$ and $\eta(t) = \int n(\xi)q^*(\xi-t)\,{\rm d}\xi$, respectively.
Here, $n(t)$ is the AWGN, represented by a complex-valued white Gaussian random process with a zero mean and a spectral density of {$\sigma_v^2$}. The correlation of $\eta(t)$ is calculated by $\mathbb{E}[\eta(aT)\eta^*(aT)]=\sigma_v^2g((a-b)T)$ \cite{IS19TWCSVD}. 

The received signal $y(t)$ is sampled with the FTN interval of $T<T_0$ to obtain the samples of 
\begin{IEEEeqnarray}{rCL}
y_n = \sum_{n^\prime=0}^{N-1} s_{n^\prime}g((n-{n^\prime})T)+\eta(nT).
\label{yn}
\end{IEEEeqnarray}
In conventional Nyquist signaling where $\tau=1$, $y(t)$ is sampled at the ISI-free interval $T_0$ so that the symbols can be demodulated unaffected by ISI, while in FTN signaling where $0<\tau<1$, $y(t)$ is sampled at the non-zero entries of $g(t)$, leading to ISI between the samples.

Furthermore, \eqref{yn} is rewritten in the vectorial form of
$\mathbf{y} = \mathbf{G}_N\mathbf{s}+\boldsymbol{\eta}$,
where 
$\mathbf{s}=[s_0,s_1,\cdots,s_{N-1}]^T {\in}\mathbb{C}^{N\times1}$, 
$\mathbf{y}=[y_0,y_1,\cdots,y_{N-1}]^T {\in}\mathbb{C}^{N\times1}$, and
$\boldsymbol{\eta} = [\eta(0),\eta(T),\eta(2T),\cdots,\eta((N-1)T)]^T  {\in}\mathbb{C}^{N\times1}$,
while 
$\mathbf{G}_N{\in}\mathbb{C}^{N{\times}N}$ is the ISI matrix, which is a $N$-order Toeplitz matrix whose first column is $\mathbf{g}=[g(0),g(T),\cdots,g((N-1)T)]^T$ and first row is $\mathbf{g}^T$. 
Noted that $\mathbf{G}_N$ can be simplified by setting an equivalent ISI length $\nu$ and approximating $g(nT)=0$ if $n>\nu$ \cite{LBZTCOM18reduced} under the assumption of $g(t)$ as a finite-impulse-response filter \cite{HSSTSP16An}.
Furthermore,
the covariance matrix of $\boldsymbol{\eta}$ is given by 
$\mathbb{E}[\boldsymbol{\eta}\boldsymbol{\eta}^H]=\sigma_v^2\mathbf{G}_N$~\cite{ISH21ACCESSThe}.

\subsection{FTN Signaling in Frequency-Selective Fading Channel}
In the frequency-selective fading channel, the FTN signaling receiver suffers from serious ISI effects due to FTN signaling and delay spread. 
At the transmitter, the CP and CS, each with a length of $\nu$, are added to every transmit signal block, where $\nu$ is set sufficiently higher than the delay spread. The transmitted symbols are formulated by
$\mathbf{s}_{CP} =\mathbf{A}_{N+2\nu}\mathbf{s}$,
where $\mathbf{A}_{N+2\nu}{\in}\mathbb{C}^{(N+2\nu){\times}N}$ is the CP/CS pending matrix with the structure~\cite{HSSTSP16An}.
The signal is sent through the frequency-selective fading channel having $L$-length tap {($L\leq\nu$)}, whose channel coefficients are $h_l$ $(l=0,1,\cdots,L-1)$ with an averaged power of $\sigma_h^2$, which is also represented by $\mathbf{h}=[h_0,\cdots,h_{L-1}]^T$. Passing through the frequency-selective fading channel and matched filter, the received samples are given by
$y_{CP,n} =\sum_{l=0}^{L-1} \sum_{k=-\nu}^{\nu} s_{CP,n-l}h_l g(nT-(l+k)T)+\eta(nT)$,
which is also expressed in the vectorized form as
$\mathbf{y}_{CP} =\boldsymbol{\Theta}_{N+2\nu} \mathbf{s}_{CP}+\boldsymbol{\eta}_{N+2\nu}$.
Here, $\boldsymbol{\Theta}_{N+2\nu}$ is the overall ISI matrix, which is expressed by the multiplication of the FTN-induced ISI matrix $\mathbf{G}_{N+2\nu}\in\mathbb{C}^{(N+2\nu){\times}(N+2\nu)}$ and the channel matrix $\mathbf{H}_{N+2\nu}\in\mathbb{C}^{(N+2\nu){\times}(N+2\nu)}$, \emph{i.e.}, $\boldsymbol{\Theta}_{N+2\nu}=\mathbf{G}_{N+2\nu}\mathbf{H}_{N+2\nu}$. More specifically, $\mathbf{G}_{N+2\nu}$ is defined in the same way as $\mathbf{G}_N$ with the equivalent ISI length of $\nu$, and the $m$th-row and $n$th-column entry of $\mathbf{H}_{N+2\nu}$ is given by $H(m,n)=h_{m-n}~(0\leq m-n\leq L-1)$ or $H(m,n)=0~(m-n > L-1)$. {$\boldsymbol{\eta}_{N+2\nu}\in\mathbb{C}^{(N+2\nu){\times}1}$ is the colored noise vector, which has a correlation matrix given by}
$\mathbb{E}[\boldsymbol{\eta}_{N+2\nu}\boldsymbol{\eta}_{N+2\nu}^H]=\sigma_v^2\mathbf{G}_{N+2\nu}$.

Then, the CP/CS-related samples are removed from $\mathbf{y}_{CP}$ to attain the CP/CS-free ones: 
\begin{IEEEeqnarray}{rCL}
\mathbf{y} &=& \mathbf{R}_{N+2\nu}\mathbf{y}_{CP} =\mathbf{R}_{N+2\nu}(\boldsymbol{\Theta}_{N+2\nu} \mathbf{s}_{CP}+\boldsymbol{\eta}_{N+2\nu}), \label{y_tdcpr}
\end{IEEEeqnarray}
where $\mathbf{R}_{N+2\nu}\in\mathbb{C}^{N{\times}(N+2\nu)}$ is the CP/CS removal matrix with the structure of $\mathbf{R}_{N+2\nu}=[\mathbf{0}_{\nu{\times}\nu},\mathbf{I}_{N},\mathbf{0}_{\nu{\times}\nu}]$ \cite{HSSTSP16An}.
The role of the CP and CS is to make the channel matrix circulant, which can be easily decomposed with the aid of the DFT. More specifically, \eqref{y_tdcpr} can be rewritten as
$\mathbf{y} =\boldsymbol{\Theta}\mathbf{s}+\boldsymbol{\eta}$,
where $\boldsymbol{\Theta}$ is an $N{\times}N$ circulant matrix, given by $\boldsymbol{\Theta}=\mathbf{G}\mathbf{H}$. Here, $\mathbf{G}$ is an $N{\times}N$ circulant matrix, whose first column is given by $[g(0),\cdots,g(\nu T), \mathbf{0}_{1{\times}(N-2\nu-1)},g(-\nu T),\cdots,g(-T)]^T$, corresponding to the circulant version of the FTN-induced ISI matrix $\mathbf{G}_N$. 
Also, $\mathbf{H}$ is an $N{\times}N$ circulant matrix, whose first column is given by $[\mathbf{h}^T,\mathbf{0}_{1{\times}(N-L)}]^T$, corresponding to the truncated and circulant version of the channel matrix $\mathbf{H}_{N+2\nu}$. The colored noise term $\boldsymbol{\eta}_{N+2\nu}$ is also converted back to $\boldsymbol{\eta}$ by removing the first and last $\nu$ entries. 
Because of the circulant property of $\boldsymbol{\Theta}$, its DFT-based decomposition is carried out with low complexity as follows:
\begin{IEEEeqnarray}{rCL}
\boldsymbol{\Theta}=\mathbf{F}_N^H \boldsymbol{\Lambda} \mathbf{F}_N,
\label{H_decomp}
\end{IEEEeqnarray}
where $\boldsymbol{\Lambda}=\text{diag}(\lambda_0,\cdots,\lambda_{N-1})$ is the eigenvalue matrix, whose diagonal entries are eigenvalues of $\boldsymbol{\Theta}$. 
Noted that for the AWGN channel, we have $\mathbf{H}=\mathbf{I}_N$, where the DFT-based decomposition of $\boldsymbol{\Theta}$ is equivalent to that of $\mathbf{G}$.

At the receiver, by carrying out the DFT of $\mathbf{y}$, we have the FD counterpart of the received samples as follows:
\begin{IEEEeqnarray}{rCL}
\tilde{\mathbf{y}} &=& \mathbf{F}_N \mathbf{y}
= \boldsymbol{\Lambda} \tilde{\mathbf{s}} + \tilde{\boldsymbol{\eta}},
\label{fde}
\end{IEEEeqnarray}
where 
$\tilde{\mathbf{s}}=\mathbf{F}_N \mathbf{s}$ and $\tilde{\boldsymbol{\eta}}=\mathbf{F}_N\boldsymbol{\eta}$.
Furthermore, 
The covariance matrix of $\tilde{\boldsymbol{\eta}}$ is given by ${\sigma_v^2}\tilde{\mathbf{G}}$, where $\tilde{\mathbf{G}}_N=\mathbf{F}_N\mathbf{G}_N\mathbf{F}_N^H$ \cite{LYYTVT20Time}.

Now, $\mathbf{s}$ is demodulated by applying the minimum mean-square error (MMSE) weighting matrix ${\mathbf{W}_{FDE}}$ and taking the {inverse discrete Fourier transform (IDFT)} on (\ref{fde}) as follows:
$\hat{\mathbf{s}} =\mathbf{F}^H {\mathbf{W}_{FDE}} \tilde{\mathbf{y}}$.
An unignorable issue here is that the noises $\tilde{\boldsymbol{\eta}}$ in the FD are correlated, where the associated covariance matrix is non-diagonal. The MMSE weighting matrix {$\mathbf{W}_{FDE}$} that whitens noises is given by
${\mathbf{W}_{FDE}} = \boldsymbol{\Lambda}^H\left(\boldsymbol{\Lambda}\boldsymbol{\Lambda}^H+\frac{{\sigma_v^2}}{\sigma_s^2}\boldsymbol{\Phi}\right)^{-1}$,
where 
$\boldsymbol{\Phi}=\text{diag}(\tilde{G}_N(0,0),\tilde{G}_N(1,1),\cdots,\tilde{G}_N(N-1,N-1))$,
and $\tilde{G}_N(n,n)$ is the $n$th diagonal element of $\tilde{\mathbf{G}_N}$~\cite{ISTWC17Iterative}.

\section{FD Pilot Superimposition for FTN Signaling}     \label{[SecCE]}

\begin{figure}
\begin{center}
\subfigure[]{
\includegraphics[width=.9\linewidth]{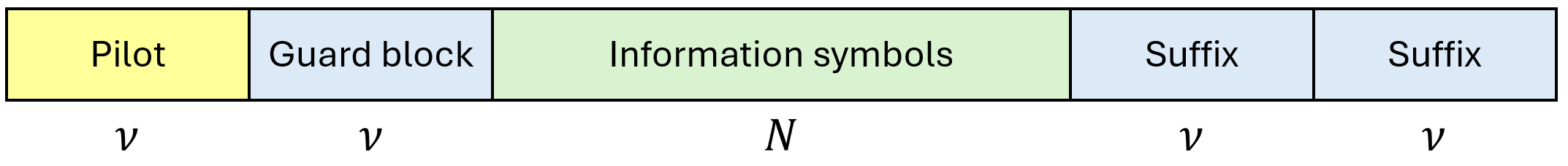}
\label{[Fig_block_bench]}
}
\subfigure[]{
\includegraphics[width=.7\linewidth]{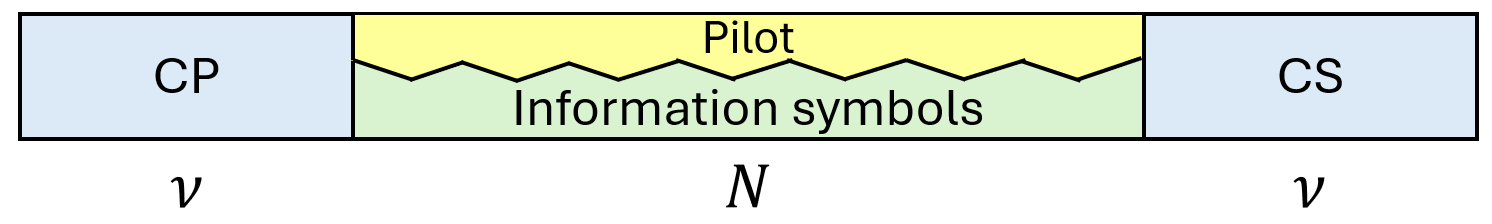}
\label{[Fig_block_proposal]}
}
\end{center}
\caption{The block structures of the benchmark CE \cite{ISTWC17Iterative} and the proposed CE based on the superimposed pilot, both for FDE-aided FTN signaling, where the total overhead of each scheme is given by $4\nu$ and $2\nu$.}
\label{[Fig_block_structure]}
\end{figure}

{Fig.~\ref{[Fig_block_structure]} shows the block structures of the conventional and the proposed CE schemes for FDE-aided FTN signaling. 
The conventional CE scheme of Fig.~\ref{[Fig_block_bench]} adopts the pilot with a length of $\nu$ symbols, and extra $3\nu$ symbols are attached as the guard block and suffix, resulting in an overhead of $4\nu$ in total. 
By contrast, in the block structure of the proposed scheme (Fig. \ref{[Fig_block_proposal]}), the overhead is a pair of CP and CS, with a length of $2\nu$ in total.}

\subsection{CE with FD Superimposed Pilot}
Denote the pilot sequence as $\mathbf{x}_p=[x_{p,0},\cdots,x_{p,N-1}]^T{\in}\mathbb{C}^{N{\times}1}$ with the power of $\sigma_p^2$ per symbol, the TD transmit symbols $\mathbf{x}_{sp}$ that superimposes information and pilot symbols are given by
\begin{IEEEeqnarray}{rCL}
\mathbf{x}_{sp} = \mathbf{s}+\mathbf{x}_p,
\label{x_sp_td}
\end{IEEEeqnarray}
and CP and CS, each having $\nu$ length, are added to $\mathbf{x}_{sp}$, making the transmitted signal vector $\mathbf{x}_{sp,cp}{\in}\mathbb{C}^{(N+2\nu){\times}1}$ become
$\mathbf{x}_{sp,cp} = \mathbf{A}_{N+2\nu}\mathbf{x}_{sp}$.

At the receiver, the TD received signal with the CP and CS is represented by
$\mathbf{y}_{sp,cp}=\boldsymbol{\Theta}_{N+2\nu}\mathbf{x}_{sp,cp}+ \boldsymbol{\eta}_{N+2\nu} {\in}\mathbb{C}^{(N+2\nu){\times}1}$.
The CP and CS are removed, and the channel matrix becomes circulant. Hence, the received samples after the CP/CS removal is given by
\begin{IEEEeqnarray}{rCL}
\mathbf{y}_{sp}&=& \mathbf{R}_{N+2\nu}(\mathbf{y}_{sp,cp}+\boldsymbol{\eta}_{N+2\nu}) {\in}\mathbb{C}^{(N+2\nu){\times}1}\\
&=&\boldsymbol{\Theta}(\mathbf{s}+\mathbf{x}_p) + \boldsymbol{\eta}.
\label{y_sp_td}
\end{IEEEeqnarray}
By taking the DFT of $\mathbf{y}_{sp}$, we can obtain its FD counterpart as follows:
\begin{IEEEeqnarray}{rCL}
\tilde{\mathbf{y}}_{sp} &=&\mathbf{F}_N \mathbf{y}_{sp}\\
&=&\boldsymbol{\Lambda}\tilde{\mathbf{s}}+\boldsymbol{\Lambda}\tilde{\mathbf{x}}_p + \tilde{\boldsymbol{\eta}},
\label{y_sp_fd}
\end{IEEEeqnarray}
where the DFT-based decomposition of $\boldsymbol{\Theta}$ is applied in the same manner as (\ref{H_decomp}) and (\ref{fde}). {$\tilde{\mathbf{x}}_p$ is the FD counterpart of ${\mathbf{x}}_p$, which is given by}
$\tilde{\mathbf{x}}_p=\mathbf{F}_N\mathbf{x}_p$.

{Recalling} that $\mathbf{G}$ and $\mathbf{H}$ are circulant matrices, the DFT-based decomposition can be carried out in the same manner as (\ref{H_decomp}). More specifically, we obtain
\begin{IEEEeqnarray}{rCL}
\boldsymbol{\Theta}=\mathbf{G}\mathbf{H}
=\mathbf{F}_N^H \boldsymbol{\Lambda_g}\boldsymbol{\Lambda_h} \mathbf{F}_N,
\label{hhg_decomp}
\end{IEEEeqnarray}
where $\boldsymbol{\Lambda_g}=\text{diag}(\lambda_{g,0},\lambda_{g,1},\cdots,\lambda_{g,N-1})$ and $\boldsymbol{\Lambda_h}=\text{diag}(\lambda_{h,0},\lambda_{h,1},\cdots,\lambda_{h,N-1})$ are the eigenvalues of $\mathbf{G}$ and $\mathbf{H}$, respectively. 
From (\ref{H_decomp}), the eigenvalues of $\boldsymbol{\Theta}$ are found to be the multiplications of the eigenvalues of $\mathbf{G}$ and $\mathbf{H}$, \emph{i.e.} $\boldsymbol{\Lambda}=\boldsymbol{\Lambda_g}\boldsymbol{\Lambda_h}$.
Hence, (\ref{y_sp_fd}) is rewritten by
\begin{IEEEeqnarray}{rCL}
\begin{aligned}
\tilde{\mathbf{y}}_{sp} = \boldsymbol{\Lambda_h}\boldsymbol{\Lambda_g}\tilde{\mathbf{s}}+\boldsymbol{\Lambda_h}\boldsymbol{\Lambda_g}\tilde{\mathbf{x}_p} + \tilde{\boldsymbol{\eta}},
\label{y_sp_fd_gh}
\end{aligned}
\end{IEEEeqnarray}
where the property of $\boldsymbol{\Lambda_g}\boldsymbol{\Lambda_h}=\boldsymbol{\Lambda_h}\boldsymbol{\Lambda_g}$ is used.
Moreover, \eqref{y_sp_fd_gh} is rewritten by
\begin{IEEEeqnarray}{rCL}
{\tilde{y}_{sp,k} = \lambda_{h,k}\lambda_{g,k} \tilde{s}_k + \lambda_{h,k}\lambda_{g,k} \tilde{x}_{p,k} + \tilde{\eta}_k,}
\label{y_sp_fd_point}
\end{IEEEeqnarray}
{where $\tilde{y}_{sp,k}$, $\tilde{s}_k$, $\tilde{x}_{p,k}$, and $\tilde{\eta}_k$ are the $k$-th entries of $\tilde{\mathbf{y}}_{sp}$, $\tilde{\mathbf{s}}$, $\tilde{\mathbf{x}_p}$, and $\tilde{\boldsymbol{\eta}}$, respectively.}

The diagonal elements $\lambda_{h,k}$ of $\boldsymbol{\Lambda_h}$ are as follows: 
$\lambda_{h,k}=\sum_{l=0}^{L-1} h_l e^{-j2{\pi}kl/N}$.
It can be found that the expression $\lambda_{h,k}$ is identical to the FD channel response of $\mathbf{h}$, \emph{i.e.}, the DFT of $[\mathbf{h}^T,\mathbf{0}_{1{\times}(N-L)}]^T$. This is because $\boldsymbol{\Lambda_h}$ is the eigenvalue matrix obtained from the DFT-based decomposition of $\mathbf{H}$, and thus it is equivalent to acquiring the FD counterpart of $\mathbf{h}$. This implies that the acquisition of $\lambda_{h,k}$ allows us to attain the channel coefficient $h_l$ via FD CE. 
Note that $\boldsymbol{\Lambda_g}$ comes from the DFT-based decomposition of $\mathbf{G}$, which can be calculated offline in advance of transmission once $\beta$ and $\tau$ are fixed. 
Hence, in our FD CE problem of (\ref{y_sp_fd_point}), $\lambda_{h,k}$ is calculated from the received sample {$\tilde{y}_{sp,k}$}, the pilot symbol {$\tilde{x}_{p,k}$} and the pre-calculated eigenvalue $\lambda_{g,k}$.

This can be simplified with the aid of a specific pilot sequence $\mathbf{x}_p$ design. More specifically, assume $\mathbf{x}_p$ consists of $Q$ periodic sequences each with the period of $P$ as $\mathbf{x}_p=[\mathbf{c}_0^T,\mathbf{c}_1^T,\cdots,\mathbf{c}_{Q-1}^T]^T$, where $P$ and $Q$ have to satisfy the relationship of $N=PQ$. 
Here, identical Chu's sequence \cite{CTIT72Poly} is adopted for $Q$ sequences as follows: $\mathbf{c}_0=\cdots=\mathbf{c}_{Q-1}{\in}\mathbb{C}^{P{\times}1}$. 
Since $\mathbf{x}_p$ consists of $Q$ repeated sequences, each being an identical Chu's sequence of period $P$, its FD counterpart $\tilde{\mathbf{x}}_p$ exhibits a highly structured pattern. The energy of $\tilde{\mathbf{x}}_p$ is concentrated only at $P$ equally spaced frequency bins, indexed by $k=iQ$ ($i=0,\cdots,P-1$), which is key to enabling FD CE in the frequency bins $k=iQ$.
Note that the relationship of $P\ge L$ has to be satisfied for stable CE.

In our algorithm, the estimated FD channel response {$\hat{\lambda}_{h,k}$} is obtained from the received samples on the partial frequency bins of $k = iQ~(i = 0,1,\cdots,P-1)$. 
Define $\tilde{\mathbf{y}}_{sp}^\prime$, $\tilde{\mathbf{x}}_p^\prime$, $\tilde{\mathbf{s}}^\prime$ and $\tilde{\boldsymbol{\eta}}^\prime$ as $P$-sized vectors, whose $i$th elements are the $(iQ)$th elements of $\tilde{\mathbf{y}}_{sp}$, $\tilde{\mathbf{x}}_p$,  $\tilde{\mathbf{s}}$ and $\tilde{\boldsymbol{\eta}}$, respectively. Then, we have
\begin{IEEEeqnarray}{rCL}
\tilde{\mathbf{y}}_{sp}^\prime = \boldsymbol{\Lambda_h}^\prime\boldsymbol{\Lambda_g}^\prime\tilde{\mathbf{s}}^\prime+\boldsymbol{\Lambda}_h^\prime\boldsymbol{\Lambda_g}^\prime\tilde{\mathbf{x}}_p^\prime + \tilde{\boldsymbol{\eta}}^\prime,
\label{y_sp_fbins}
\end{IEEEeqnarray}
where $\boldsymbol{\Lambda_h}^\prime$ and $\boldsymbol{\Lambda_g}^\prime$ are $P{\times}P$ diagonal matrices, whose $i$th diagonal elements are the $(iQ)$th diagonal element of $\boldsymbol{\Lambda_h}$ and $\boldsymbol{\Lambda_g}$.

By regarding the interference term of $\boldsymbol{\Lambda_h}^\prime\boldsymbol{\Lambda_g}^\prime\tilde{\mathbf{s}}^\prime$ in \eqref{y_sp_fbins} as the noise, the FD channel response is estimated as
\begin{IEEEeqnarray}{rCL}
\mathbf{\hat{d}}&=& [{\hat{\lambda}_{h,0},\hat{\lambda}_{h,Q},\hat{\lambda}_{h,2Q},\cdots,\hat{\lambda}_{h,(P-1)Q}}]^T
=\mathbf{W}\tilde{\mathbf{y}}_{sp}^\prime,
\label{d_est}
\end{IEEEeqnarray}
where $\mathbf{W}$ is the $P{\times}P$-sized complex-valued  weight matrix. 
For example, we can design the weights for the least-square (LS) estimation as 
\begin{IEEEeqnarray}{rCL}
\mathbf{W}_{LS}=\boldsymbol{\Gamma}^{-1},
\label{wls}
\end{IEEEeqnarray}
where
$\boldsymbol{\Gamma}=\text{diag}(\gamma_0,\gamma_1,\cdots,\gamma_{P-1})$ and
$\gamma_{i}={\Lambda}_{g,i}^\prime\cdot\tilde{{x}}_{p,i}^\prime$.
{Note that ${\Lambda}_{g,i}^\prime$ and $\tilde{{x}}_{p,i}^\prime$ are the $i$-th entries of $\boldsymbol{\Lambda_g}^\prime$ and $\tilde{\mathbf{x}}_p^\prime$, respectively.}

Furthermore, the weights for the MMSE-based estimator are formulated by
\begin{IEEEeqnarray}{rCL}
\mathbf{W}_{MMSE}
&=&\boldsymbol{\Gamma}^H(\boldsymbol{\Gamma}\boldsymbol{\Gamma}^H+\frac{\sigma_{intf}^2}{\sigma_h^2P}\boldsymbol{\Phi}_{intf})^{-1},
\label{wmmse}
\end{IEEEeqnarray}
where {$\mathbf{R}_d$ is the correlation matrix of $\mathbf{d}=[{\lambda}_{h,0},{\lambda}_{h,Q},\cdots,{\lambda}_{h,(P-1)Q}]^T$ under the ideal case that is free of interference, expressed as}
$\mathbf{R}_d={\mathbb{E}[\mathbf{d}\mathbf{d}^H]=P\mathbb{E}[\mathbf{h}\mathbf{h}^H]}=\sigma_h^2P\mathbf{I}_P$.
Here, {$\sigma_{intf}^2$ is the variance of the sum of terms $\boldsymbol{\Lambda_h}^\prime\boldsymbol{\Lambda_g}^\prime\tilde{\mathbf{s}}^\prime+ \tilde{\boldsymbol{\eta}}^\prime$. Also, $\boldsymbol{\Phi}_{intf}$ denotes the correlation matrix of interference, including both the colored noise $\tilde{\boldsymbol{\eta}}^\prime\in\mathbb{C}^{P{\times}P}$ and the information symbol term $\boldsymbol{\Lambda_h}^\prime\boldsymbol{\Lambda_g}^\prime\tilde{\mathbf{s}}^\prime$, given by}
$\boldsymbol{\Phi}_{intf}
=\sigma_s^2\mathbb{E}[\boldsymbol{\Lambda_h}^\prime\boldsymbol{\Lambda_g}^\prime(\boldsymbol{\Lambda_g}^\prime)^H(\boldsymbol{\Lambda_h}^\prime)^H]+\mathbb{E}[\tilde{\boldsymbol{\eta}}^\prime(\tilde{\boldsymbol{\eta}}^\prime)^H]$,
where the properties $\mathbb{E}[\tilde{s}^\prime(\tilde{s}^\prime)^H]=\sigma_s^2\mathbf{I}_P$ and $\mathbb{E}[\tilde{s}^\prime\tilde{\boldsymbol{\eta}}^\prime]=\mathbf{0}_{P{\times}P}$ are used. Note that $\boldsymbol{\Phi}_{intf}$ is not diagonal. Moreover, the existence of the term $\boldsymbol{\Lambda_h}^\prime\boldsymbol{\Lambda_g}^\prime\tilde{\mathbf{s}}^\prime$ makes it challenging to calculate $\mathbf{W}_{MMSE}$.

Finally, the TD channel coefficients are estimated by carrying out the IDFT of $\mathbf{\hat{d}}$ with the power scaling $1/\sqrt{P}$ as follows:
\begin{IEEEeqnarray}{rCL}
\mathbf{\hat{h}} = \frac{1}{\sqrt{P}} \mathbf{F}^H_{P,L} \mathbf{\hat{d}}.
\label{h_td_est}
\end{IEEEeqnarray}
Under the assumption of $P \ge L$, 
the TD channel matrix $\hat{\mathbf{H}}$ can be reconstructed as the $N{\times}N$ circulant matrix with $[\hat{\mathbf{h}}^T,\mathbf{0}_{1{\times}(N-L)}]^{T}$ as the first column.

\subsection{Interference-Free CE by {SIA}}

In order to minimize the above-mentioned interference in CE, we invoke the {SIA} concept~\cite{GMALSP05Channel} for the proposed scheme.
The main idea is to subtract $\mathbf{s}$ with the aid of a specially-designed data-dependent sequence, so that the frequency bins of the index $k=iQ$ can align with the zero point on the spectrum, and $\hat{\lambda}_{h,iQ}$ can be calculated affected only by the noise term $\tilde{\boldsymbol{\eta}}^\prime$, rather than the interference term $\boldsymbol{\Lambda_h}^\prime\boldsymbol{\Lambda_g}^\prime\tilde{\mathbf{s}}^\prime$.

First, let us define the {data-dependent sequence} as $\boldsymbol{\Upsilon}=[\Upsilon_0,\Upsilon_1,\cdots,\Upsilon_{N-1}]^T{\in}\mathbb{C}^{N{\times}1}$ and the FD counterpart as $\tilde{\boldsymbol{\Upsilon}}={\mathbf{F}_N\boldsymbol{\Upsilon}=}[\tilde\Upsilon_0,\tilde\Upsilon_1,\cdots,\tilde\Upsilon_{N-1}]^T$, where
\begin{IEEEeqnarray}{rCL}	
{\tilde{\Upsilon}_k}=
\begin{cases}
{\tilde{s}_k}, \ \ \textrm{for} \  k=iQ\\
0, \ \ \text{otherwise}
\end{cases}.
\label{j_fd}
\end{IEEEeqnarray}
This allows us to have the following relationship:
\begin{IEEEeqnarray}{rCL}
{\tilde{s}_k-\tilde{\Upsilon}_k}=
\begin{cases}
0, \ \ \textrm{for} \  k=iQ\\
{\tilde{s}_k}, \ \ \text{otherwise}
\end{cases}.
\label{upsilon_fd}
\end{IEEEeqnarray}
Hence, the DFT of $\mathbf{s}-\boldsymbol{\Upsilon}$ corresponds to zeros on the frequency bins $k=iQ$, which implies that interference associated with information symbols may be removed.

The {data-dependent sequence} $\boldsymbol{\Upsilon}$ has the cyclic property as follows: 
${\Upsilon_{i+mP}} = {\Upsilon_i} 
 = \frac{1}{Q}\sum_{m=0}^{Q-1} {s_{i+mP}} \ \ \textrm{for}~ i = 0,\cdots,P-1$.
{With the aid of SIA, the transmitted symbol vector $\mathbf{x}_{sp,d}$ is obtained by subtracting the data-dependent sequence, \emph{i.e.}, the cyclic means of $\mathbf{s}$}, which are represented, instead of (\ref{x_sp_td}), by \cite{GMALSP05Channel}
\begin{IEEEeqnarray}{rCL}
\mathbf{x}_{sp,d} = (\mathbf{I}-\mathbf{J})\mathbf{s}+\mathbf{x}_p,
\label{x_sp_td_j}
\end{IEEEeqnarray}
where 
$\mathbf{J} = \frac{1}{Q} \mathbf{1}_Q {\otimes} \mathbf{I}_P$.

{The SIA-processed information symbols} $(\mathbf{I}-\mathbf{J})\mathbf{s}$ in (\ref{x_sp_td_j}) exhibit a total power reduction from the conventional information symbols $\mathbf{s}$ by the factor of $1/Q$. 
In order to compensate for this, the pilot power $\sigma_p^2$ is adjusted. In this paper, we assume the total transmit power per symbol remains unchanged at $\sigma_s^2$ either with or without {SIA}, where the pilot power per symbol is set to be $\sigma_p^2=(1-1/Q)\sigma_s^2$ for the {SIA} scheme. 

Finally, by transmitting $\mathbf{x}_{sp,d}$ of (\ref{x_sp_td_j}), the FD received samples in (\ref{y_sp_fbins}) becomes
\begin{IEEEeqnarray}{rCL}
\tilde{\mathbf{y}}_{sp,d}^\prime = \boldsymbol{\Lambda_h}^\prime\boldsymbol{\Lambda_g}^\prime\tilde{\mathbf{x}}_p^\prime + \tilde{\boldsymbol{\eta}}^\prime,
\label{y_sp_fbins_dr}
\end{IEEEeqnarray}
from which the frequency channel response $\boldsymbol{\Lambda_h}^\prime$ can be calculated without interference associated with the information symbols, according to (\ref{d_est})--(\ref{h_td_est}). 
Moreover, owing to the elimination of $\mathbf{s}$ in \eqref{y_sp_fbins_dr}, the MMSE weights are simplified to
$\mathbf{W}_{MMSE,d}=\mathbf{R}_d\left(\boldsymbol{\Gamma}\boldsymbol{\Gamma}^H\mathbf{R}_d+\sigma_{v}^2\boldsymbol{\Phi}^{\prime}\right)^{-1}\boldsymbol{\Gamma}^H
=\boldsymbol{\Gamma}^H\left(\boldsymbol{\Gamma}\boldsymbol{\Gamma}^H+\frac{\sigma_{v}^2}{\sigma_{h}^2}\boldsymbol{\Phi}^{\prime}\right)^{-1}$,
where
${\boldsymbol{\Phi}}^{\prime}=\text{diag}(\Phi[0],\Phi[Q],\cdots,\Phi[(P-1)Q])$,
which is the diagonal approximation of $\mathbb{E}[\tilde{\boldsymbol{\eta}}^{\prime}(\tilde{\boldsymbol{\eta}}^{\prime})^H]$. The LS weights $\mathbf{W}_{LS}$ remain the same as \eqref{wls}.

\subsection{FDE}
{We now calculate the FD channel response $\hat{\lambda}^{eq}_{h,k}$, which is used for equalization as follows: }
${\hat{\lambda}^{eq}_{h,k}} = \sum_{l=0}^{L-1} \hat{h_l}e^{-j2{\pi}kl/N}$,
and thus we obtain $\hat{\boldsymbol{\Lambda}}_{h} = \textrm{diag}({\hat\lambda^{eq}_{h,0}, \cdots,\hat\lambda^{eq}_{h,N-1}})$. The information symbols are equalized with FDE as follows:
\begin{IEEEeqnarray}{rCL}
\mathbf{u} = \mathbf{F}^H_N\mathbf{W}_{eq}\tilde{\mathbf{z}},
\label{u_fd}
\end{IEEEeqnarray}
where the $k$th entry of $\tilde{\mathbf{z}}$, {denoted as $\tilde{{z}}_k$} is given by 
\begin{IEEEeqnarray}{rCL}
{\tilde{{z}}_k}=
\begin{cases}
0, \ \ \textrm{for} \ k=iQ\\
{\tilde{{y}}_{sp,k}}, \ \ \text{otherwise}
\end{cases}.
\label{z_fd}
\end{IEEEeqnarray}
In \eqref{z_fd}, the symbols on the frequency bins used for CE are removed.
Also, $\mathbf{W}_{eq}$ is the equalization weights. More specifically, the LS and MMSE weights are represented, respectively, by 
$\mathbf{W}_{eq,LS} = \boldsymbol{\Gamma}_{eq}^{-1}$ and
$\mathbf{W}_{eq,MMSE} = \boldsymbol{\Gamma}_{eq}^{H}\left(\boldsymbol{\Gamma}_{eq}\boldsymbol{\Gamma}_{eq}^{H}+\frac{\hat{\sigma_v}^2}{\hat{\sigma_s}^2}\boldsymbol{\Phi}\right)^{-1}$,
where 
$\boldsymbol{\Gamma}_{eq}=\hat{\boldsymbol{\Lambda}}_{h}\boldsymbol\Lambda_{g}
=\text{diag}(\gamma_{eq,0},\cdots,\gamma_{eq,N-1})$,
$\hat{\sigma_s}^2=(1-1/Q)\sigma_s^2$, and
$\hat{\sigma_v}^2=(1-1/Q)\sigma_v^2$.
Owing to the diagonal structures of $\boldsymbol{\Gamma}_{eq}$ and $\boldsymbol{\Phi}$, (\ref{u_fd}) can be calculated with low complexity in a symbol-by-symbol manner.

\subsection{Symbol Detection}

The equalized symbols $\mathbf{u}$ in \eqref{u_fd} correspond to the estimates of the {SIA-processed information symbols} $(\mathbf{I}-\mathbf{J})\mathbf{s}$ in (\ref{x_sp_td_j}), which, hence, can not be directly used for symbol detection as they are. However, since $(\mathbf{I}-\mathbf{J})$ is a singular matrix, it is a challenging task to calculate its inverse matrix. Alternatively, we employ the Moore-Penrose pseudo inverse of $(\mathbf{I}-\mathbf{J})$. 

More specifically, by denoting the Moore-Penrose pseudo-inverse of $\boldsymbol{\Psi}=\mathbf{I}-\mathbf{J}$ as $\boldsymbol{\Psi}^{+}$, the initial estimates of information symbols $\mathbf{s}$ are calculated as
$\hat{\mathbf{s}}^{(0)} = \boldsymbol{\Psi}^{+}\mathbf{u}$.
Note that $\boldsymbol{\Psi}$ is an idempotent matrix, which satisfies $\boldsymbol{\Psi}^2=\boldsymbol{\Psi}$, where the Moore-Penrose pseudo-inverse of $\boldsymbol{\Psi}$ is also the matrix itself, \emph{i.e.}, $\boldsymbol{\Psi}^{+}= \boldsymbol{\Psi}$. 
Hence, we can calculate $\hat{\mathbf{s}}^{(0)}$ directly by $\boldsymbol{\Psi}\mathbf{u}$ without making extra efforts to calculate $\boldsymbol{\Psi}^{+}$.

The information symbols ${\mathbf{s}}$ are iteratively detected based on the iterative soft threshold algorithm (ISTA) as follows:~\cite{WNFTSP09Sparse}
\begin{IEEEeqnarray}{rCL}
\label{eq:1}
\hat{\mathbf{s}}^{(i+1)} &=& 
\begin{matrix}
\quad  \\
\text{argmin}\\
\mathbf{c}
\end{matrix}
\left\|\left(\hat{\mathbf{s}}^{(i)}+\boldsymbol{\Psi}^T\mathbf{r}^{(i)}\right)-\mathbf{c}\right\|^2\\
\mathbf{r}^{(i)} &=& \mathbf{u} - \boldsymbol{\Psi}\hat{\mathbf{s}}^{(i)}
\label{s_ite},
\end{IEEEeqnarray}
where each element of $\mathbf{c}$ is selected from the constellation employed for modulation. 
$\hat{\mathbf{s}}^{(i+1)}$ of \eqref{eq:1} and $\mathbf{r}^{(i)}$ of \eqref{s_ite} are calculated in an iterative manner, where each equation can be updated. This ISTA-like solution recovers $\mathbf{s}$ with the complexity as low as $\mathcal{O}(N)$ per iteration. 

\section{Performance Analysis}      \label{[SecAnalysis]}

Here, we derive the MSE of the proposed LS and MMSE CE schemes based on the FD pilot superimposition. For the LS-based CE with the {SIA}, the MSE is given by
$\sigma_{\hat{h},LS}
=\frac{L\sigma_v^2}{P^2} \text{Tr}\left\{\boldsymbol{\Gamma}^{-1}\boldsymbol{\Phi}^{\prime}(\boldsymbol{\Gamma}^{-1})^H\right\}$,
where $\mathbb{E}[\mathbf{F}^H_{P,L}\mathbf{F}_{P,L}]=(L/P)\mathbf{I}_{P}$ and $\mathbb{E}[\tilde{\boldsymbol{\eta}}^{\prime}(\tilde{\boldsymbol{\eta}}^{\prime})^H]=\sigma_v^2\boldsymbol{\Phi}^{\prime}$.
Owing to the employment of {SIA}, the information symbols $\mathbf{s}$ no longer affect the MSEs. 

Furthermore, for the MMSE-based CE, the MSE is formulated by
\begin{IEEEeqnarray}{rCL}
\sigma_{\hat{h},MMSE}&=&
\mathbb{E}\left[\left\|\mathbf{h}-\frac{1}{\sqrt{P}} \mathbf{F}^H_{P,L} \mathbf{W}_{MMSE,d}\tilde{\mathbf{y}}_{sp,d}^\prime\right\|^2\right]. \ \ \
\label{mse_mmse_1}
\end{IEEEeqnarray}
For convenience, we may calculate the MSE in terms of the FD response $\mathbf{d}$ as follows: 
\begin{IEEEeqnarray}{rCL}
&&\sigma_{\hat{h},MMSE}
\approx\frac{L}{P^2}\text{Tr}\{\mathbf{R}_d -\mathbf{R}_d\boldsymbol{\Gamma}\mathbf{W}_{MMSE,d}^H- \nonumber\\
&&\mathbf{W}_{MMSE,d}\boldsymbol{\Gamma}\mathbf{R}_d 
\left.+\mathbf{W}_{MMSE,d}(\boldsymbol{\Gamma}\mathbf{R}_d\boldsymbol{\Gamma}^H+\boldsymbol{\Phi}^{\prime})\mathbf{W}_{MMSE,d}^H\right\}, \nonumber
\end{IEEEeqnarray}
where $\mathbb{E}\left[\mathbf{d}(\tilde{\mathbf{y}}_{sp,d}^{\prime})^H\right]=\boldsymbol{\Gamma}\mathbf{R}_d$, $\mathbb{E}\left[\tilde{\mathbf{y}}_{sp,d}^{\prime}\mathbf{d}^H\right]=\mathbf{R}_d\boldsymbol{\Gamma}$, and $\mathbb{E}\left[\tilde{\mathbf{y}}_{sp,d}^{\prime}(\tilde{\mathbf{y}}_{sp,d}^{\prime})^H\right]=\boldsymbol{\Gamma}\mathbf{R}_d\boldsymbol{\Gamma}^H+\boldsymbol{\Phi}^{\prime}$ \cite{ISTWC17Iterative}.

\section{Simulation Results}        \label{[SecResults]}

In this section, we provide our performance results of the proposed CE and detection scheme in terms of the MSE and BER, where the conventional benchmark is FDE-aided FTN signaling~\cite{ISTWC17Iterative}. $E_b/N_0$ is defined by $\sigma_s^2/\sigma_v^2~(\text{dB}) = E_b/N_0~(\text{dB}) + 10\text{log}_{10}(SE)$, where $SE$ is the spectrum efficiency. The tap length of the frequency-selective fading channel is assumed to be $L=8$, and the associated power is normalized to $\sum|h_l|^2=1$ and $\sigma_h^2=1/L$, where we have $\mathbf{R}_d=(P/L)\mathbf{I}_P$. Chu's sequence~\cite{CTIT72Poly} with $(P,Q)=(8,16)$ is employed for the pilot symbols. The number of iterations employed for ISTA-based detection is set to three. The iterative joint CE and detection of the benchmark~\cite{ISTWC17Iterative} is deactivated for fair comparisons under a similar complexity. Here, quadrature phase-shift keying (QPSK) is adopted for the modulation. The MMSE criterion is used for CE and FDE unless otherwise noted. The roll-off factor of the RRC shaping filter is maintained at $\beta=0.5$, and the packing ratio is given by $0.8$ by default, though different values will also be adopted when evaluating the effect of $\tau$. The short block length is considered, where we have $N=128$, and the lengths of the CP and CS are $\nu=10$.

\begin{figure}
\begin{center}
\subfigure[]{
\includegraphics[width=.7\linewidth]{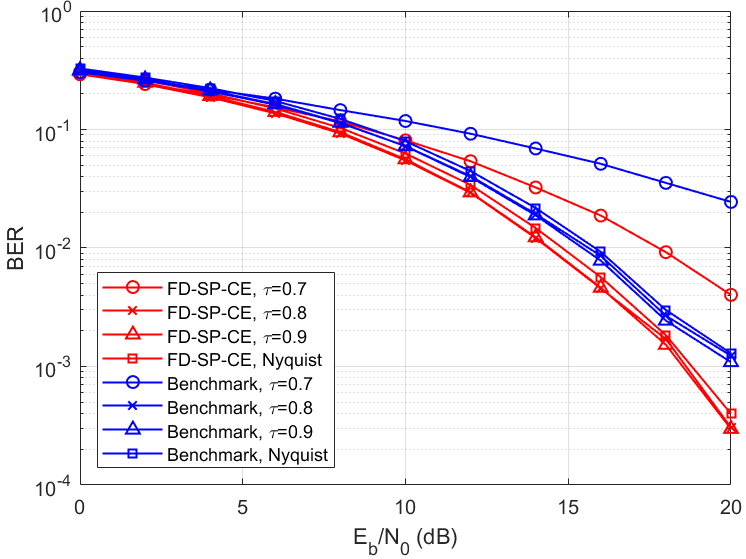}
\label{[Fig_BER_tau]}
}
\subfigure[]{
\includegraphics[width=.7\linewidth]{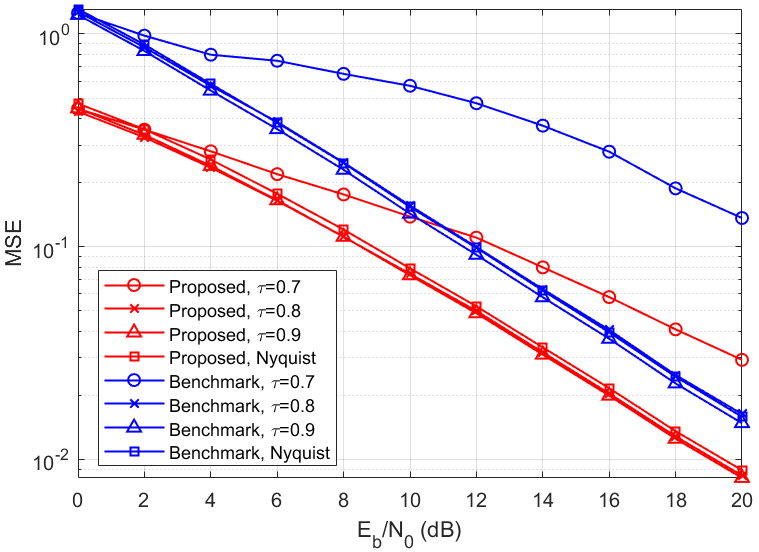}
\label{[Fig_MSE_tau]}
}
\end{center}
\caption{The BER and MSE performance of the proposed and benchmark schemes for the packing ratio $\tau=0.7$, $0.8$, and $0.9$: (a) BER and (b) MSE.}
\label{[Fig_simu_tau]}
\end{figure}

Figs. \ref{[Fig_BER_tau]} and \ref{[Fig_MSE_tau]} show the BER and MSE performance of the proposed and benchmark schemes, respectively. It can be found that the proposed scheme outperforms the benchmark in terms of BERs and MSEs for each $\tau$ value. Also, for $\tau=0.8$, $0.9$, and $1$ (i.e., the Nyquist case), the achievable BER and MSE performance remains almost the same, while the explicit deterioration is found for $\tau=0.7$. Note that the SE of Nyquist is lower than that of the FTN signaling scenarios. More specifically, the proposed scheme's BER advantage in Fig. \ref{[Fig_BER_tau]} comes from the improved SE owing to our pilot superimposition. Also, the MSE of the proposed scheme in Fig. \ref{[Fig_MSE_tau]} is achieved by our {SIA} that reduces interference of the pilot superimposition.

\begin{figure} [t]
\centering 
\includegraphics[width=0.7\linewidth]{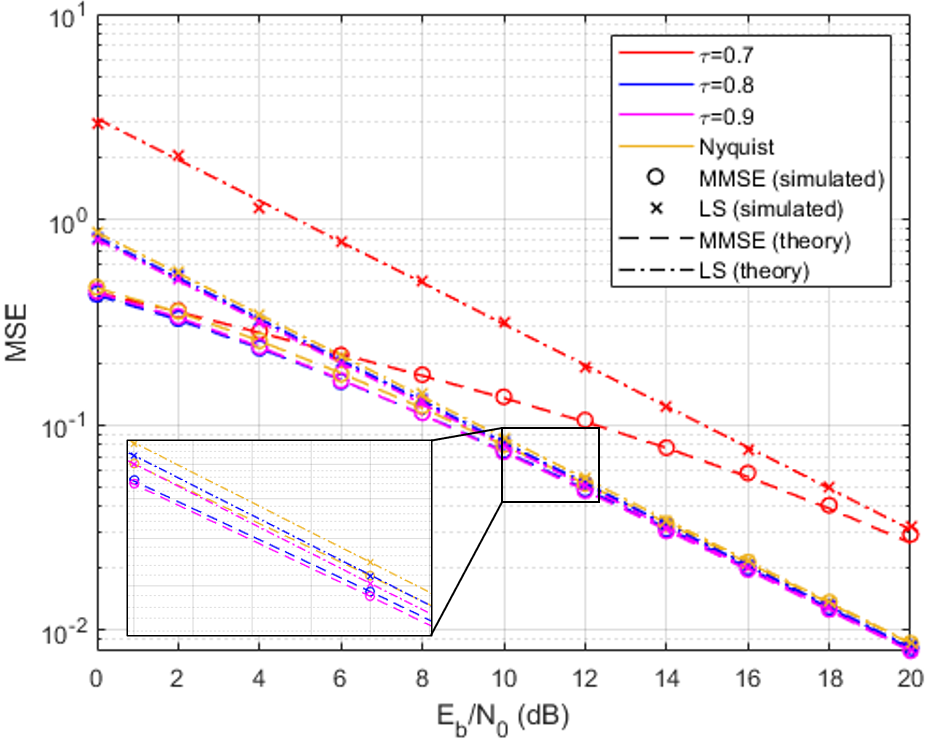}
\caption{The theoretical and numerical MSE comparisons, where we consider the LS and MMSE CE schemes for the packing ratio $\tau=0.7$, $0.8$, and $0.9$.} 
\label{[Fig_mse]}
\end{figure}

Fig. \ref{[Fig_mse]} shows the theoretical and the simulated MSEs of the proposed CE schemes based on the MMSE and LS algorithms for the packing ratios of $\tau=0.7$, $0.8$, and $0.9$. Firstly, the theoretical and numerical results match well, which verifies our system model and MSE derivation, except for the severe $\tau=0.7$ scenario. As expected, the MMSE-based CE outperforms the LS-based counterpart, while the performance gap decreases upon increasing the SNRs. This is because the MMSE algorithm takes into account the effects of the AWGN, {along with the FTN-specific colored noise.}

\section{Conclusions}  \label{[SecConclusion]}
In this paper, we proposed the novel reduced-complexity and reduced-overhead FD CE scheme for FDE-aided FTN signaling, based on the concepts of pilot superimposition and {SIA}.
While pilot superimposition dispenses with the pilot overhead, {SIA} allows us to reduce the effects of interference induced by pilot superimposition. We also derived the theoretical MSE bound for performance analysis. Our simulation results demonstrated that the proposed scheme outperforms the conventional CE benchmark for the FTN signaling in terms of the BER and MSE.
\section*{Acknowledgement}
This work was supported in part by JST SPRING (Grant JPMJSP2108), in part by JST FOREST (Grant JPMJFR2127), in part by JST ASPIRE (Grant JPMJAP2345), and in part by JSPS KAKENHI (Grant 23K22752).

\bibliographystyle{IEEEtran}
\bibliography{RefTex}

\end{document}